\documentclass[twocolumn,twoside,preprintnumbers,amsmath,amssymb,showkeys]{revtex4}
 
\usepackage{epsfig}
 
\usepackage{graphicx}% Include figure files
 
\usepackage{fancyhdr}
\usepackage{pslatex}
 
\newcommand{\lthreepaper}{$HOME/l3/paper/}
\newcommand{\lthreebiblio}{\lthreepaper/biblio/}
\newcommand{\mydirfig}{figs/}
\usepackage{physics}
\usepackage{dcolumn}
\usepackage{pennames}
\usepackage{subscript}
\usepackage{rotating}
 
\newlength{\figwidth}
\newcommand{\PZ}{\ensuremath{\mathrm{Z}}}%      Z not in pennames
\newcommand{\JETSET}{{\scshape Jetset}}

\newcommand{\Lthree}{{\scshape l}{\small 3}}

\newcommand{\beqa}{\begin{eqnarray}}   \newcommand{\eeqa}{\end{eqnarray}}
\newcommand{\beqan}{\begin{eqnarray*}} \newcommand{\eeqan}{\end{eqnarray*}}

\newcommand{\ycut}{\ensuremath{y_\mathrm{cut}}}%           % ycut
\renewcommand{\pt}{\ensuremath{p_\mathrm{t}}}%             % pt
\newcommand{\mt}{\ensuremath{m_\mathrm{t}}}%               % mt

\newcommand{\chisq}{\ensuremath{\chi^2}}%                  % chisq
 
\newcommand{\Eq}[1]{Eq.\;(\ref{#1})}%
\newcommand{\Eqs}[1]{Eqs.\;(\ref{#1})}%
\newcommand{\Fig}[1]{Fig.\;\ref{#1}}%
\newcommand{\Figs}[1]{Figs.\;\ref{#1}}%
\newcommand{\Tab}[1]{Table~\ref{#1}}%
\renewcommand{\ee}{\ensuremath{\mathrm{e^{+}e^{-}}}}

\newcommand{\Letter}{article}
\begin{document}
\title{Parametrization of Bose-Einstein Correlations and Reconstruction of the Source Function in Hadronic
Z-boson Decays using the L3 Detector\footnote{Invited talk at
\textit{II Workshop on Particle Correlation and Femtoscopy, S\~ao Paulo, 9--11 Nov., 2006,}
presented by W.J. Metzger}         }
 
\author{
        T. Nov\'ak,
        T. Cs\"org\H{o},\footnote{Visitor from   Budapest, Hungary,
             sponsored by the Scientific Exchange between Hungary (OTKA) and The Netherlands (NWO),
             project B64-27/N25186.}
        W. Kittel
        and
        W.J. Metzger %  \footnote{Presented by W.J. Metzger}
        \\ for the L3 Collaboration
       }
 
\affiliation{Radboud University, Toernooiveld 1, 6525 ED\ \ Nijmegen, The Netherlands}

\begin{abstract}
Bose-Einstein correlations of pairs of identical charged pions
produced in hadronic Z decays are analyzed in terms of various parametrizations.
A good description is achieved using a L\'evy stable distribution in conjunction with a
hadronization model having highly correlated configuration and momentum space, the $\tau$-model.
Using these results, the source function is reconstructed.
 
\end{abstract}
 
\vskip -1.35cm
 
\maketitle
 
\thispagestyle{fancy}
 
\setcounter{page}{1}

\section{Introduction}\label{sect:intr}
In particle and nuclear physics, intensity interferometry provides a direct
experimental method for the determination of sizes, shapes and lifetimes
of particle-emitting sources
(for reviews see ~\cite{Gyulassy:1979,Boal:1990,Baym:1998,Wolfram:Zako2001,Tamas:HIP2002}).
In particular, boson interferometry provides a powerful tool for the
investigation of the space-time structure of particle production processes,
since Bose-Einstein correlations (BEC) of two identical bosons reflect both
geometrical and dynamical properties of the particle radiating source.
 
%In $\mathrm{e}^+\mathrm{e}^-$ annihilation BEC are maximal when the invariant
%momentum difference, $Q$, is small, even when one of the relative momentum components
%is large, as has been previously observed\cite{TASSO:1986}.
%This is not the case in hadron-hadron interactions \cite{NA22q}
%or in heavy-ion interactions \cite{Solz:thesis,Solz:paper},
%where BEC are found not to depend simply on $Q$, but to
%decrease when any of the relative momentum components is large,
%a behavior that can be described by hydrodynamical models of the source
%\cite{Tamas;Lorstad:1996,Tamas:HIP2002}.
 
Here we study BEC in hadronic \PZ\ decay.
We investigate various static parametrizations in terms of the four-momentum difference,
$Q=\sqrt{-(p_1-p_2)^2}$
and find that none give an adequate description of the Bose-Einstein correlation function.
However, within the framework of models assuming
strongly correlated coordinate and momentum space a good description is achieved.
We then reconstruct the complete
space-time picture of the particle emitting source in hadronic Z decay.

%\section{Data} \label{sect:data}
%\subsection{Event and Charged-Particle Selection}
The data used in the analysis were collected by the
\Lthree\ detector~\cite{l3:construction,L3:Ecal_calib,L3:Hcal,L3:TEC,l3:SMD}
at an \Pep\Pem\    center-of-mass energy of $\sqrt{s}\simeq 91.2$ \GeV.
Approximately 36 million like-sign pairs of well-measured charged tracks of about 0.8 million
hadronic Z decays are used \cite{tamas}.
 
We perform analyses on the complete sample as well as  on two- and three-jet
samples. The latter are found using calorimeter clusters with the Durham jet
algorithm~\cite{durham,durham2,durham3}    % ~\cite{durham,durham1,durham2,durham3}
with a jet resolution parameter $\ycut=0.006$.
To determine the thrust axis of the event we also use calorimeter clusters.

%\section{Analysis}                     \label{sect:anal}
%\subsection{Bose-Einstein Correlation Function}
\section{Bose-Einstein Correlation Function}
The two-particle correlation function of two particles with
four-momenta $p_{1}$ and $p_{2}$ is given by the ratio of the two-particle number density,
$\rho_2(p_{1},p_{2})$,
to the product of the two single-particle number densities, $\rho_1 (p_{1})\rho_1 (p_{2})$.
Since we are here interested only in the correlation $R_2$ due to Bose-Einstein
interference, the product of single-particle densities is replaced by
$\rho_0(p_1,p_2)$,
the two-particle density that would occur in the absence of Bose-Einstein correlations:
\begin{equation} \label{eq:R2def}
  R_2(p_1,p_2)=\frac{\rho_2(p_1,p_2)}{\rho_0(p_1,p_2)} \;.
\end{equation}
This $\rho_2$ is corrected for detector acceptance and efficiency using Monte Carlo events,
to which a full detector simulation has been applied, on a bin-by-bin basis.
An event mixing technique is used to construct $\rho_0$.  This technique removes all
correlations, not just Bose-Einstein. Hence, $\rho_0$  is corrected for this using Monte Carlo~\cite{tamas}.
 
Since the mass of the two identical particles of the pair is fixed to the pion mass,
the correlation function is defined in six-dimensional momentum space.
Since Bose-Einstein correlations can be large only at small four-momentum difference
$Q$,
%$Q=\sqrt{-(p_1-p_2)^2}$,
they are often
parametrized in this one-dimensional distance measure. There is no reason, however,
to expect the hadron source to be spherically symmetric in jet fragmentation.
Recent investigations have, in fact, found an elongation of the source along the
jet axis~\cite{L3_3D:1999,OPAL3D:2000,DELPHI2D:2000,ALEPH:2004}.
While this effect is      well established, the elongation is actually only about 20\%,
which suggests that a parametrization in terms of the single variable $Q$,
may be a good approximation.
 
\begin{figure}
  \centering
  \includegraphics[width=.75\figwidth]{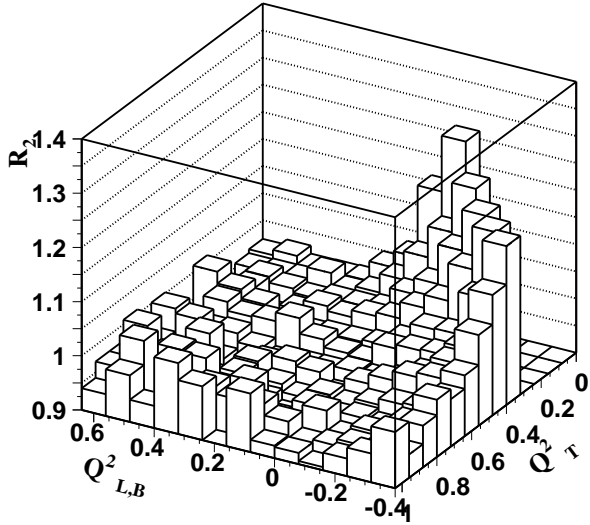}
   \caption{ $R_2$ for two-jet events as function of the squares of the transverse momentum difference
                and the combination of longitudinal momentum difference and energy difference.
%  \caption{For two-jet events:
%           (a) $R_2$ as function of the squares of the 3-momentum difference, $q$,
%               and the energy difference, $q_0$;
%           (b) $R_2$ \vs\ $q^2$ when $q^2\approx q^2_0$;
%      and  (c) $R_2$ as function of the squares of the transverse momentum difference
%               and the combination of longitudinal momentum difference and energy difference.
           \label{fig:qinv}
%\enote{Need units on the axes. $R$ should be $R_2$.}
           }
\end{figure}

This is not the case in heavy-ion and hadron-hadron interactions, where BEC are found not to depend simply
on $Q$, but on components of the momentum difference separately
\cite{Tamas:HIP2002, Tamas;Lorstad:1996, NA22q,Solz:thesis,Solz:paper}.
However, in \Pep\Pem\ annihilation at lower energy~\cite{TASSO:1986} it has been observed that $Q$ is the
appropriate variable.
We checked this and confirm that this is indeed the case:
We observe \cite{tamas} that $R_2$ does not decrease when both $q^2=(\vec{p}_1-\vec{p}_2)^2$ and
$q_0^2=(E_1-E_2)^2$ are large while $Q^2=q^2-q_0^2$ is small, but is maximal for
$Q^2=q^2-q_0^2=0$,
independent of the individual values of $q$ and $q_0$.
The same is observed in a different decomposition: $Q^2=Q_\mathrm{t}^2 + Q_{L,B}^2$, where
$Q_\mathrm{t}^2=(\vec{p}_\mathrm{t1}-\vec{p}_\mathrm{t2})^2$ is the component transverse to the thrust axis
and
$Q_{L,B}^2=(p_\mathrm{l1}-p_\mathrm{l2})^2-(E_1-E_2)^2$ combines the longitudinal momentum and energy differences.
Again, $R_2$ is maximal along the line $Q=0$, as is shown in \Fig{fig:qinv}.
This is observed both for two-jet and three-jet events.
We conclude that a parametrization in terms of  $Q$
can be considered a good approximation for the purposes of this \Letter.

\section{Parametrizations of BEC}              \label{sect:param}
With a few assumptions~\cite{GGLP:1960,Boal:1990,Tamas:HIP2002},
the two-particle correlation function, \Eq{eq:R2def},
is related to the Fourier transformed source distribution:
\begin{equation}  \label{eq:R2fourier}
     R_2(p_1,p_2) = \gamma \left[ 1 + \lambda |\tilde{f} (Q)|^2 \right]
                    \left(1 + \delta Q \right) \;,
\end{equation}
where $f(x)$ is the (configuration space) density distribution of the source,
% $Q$ is the invariant four-momentum difference, $Q=\sqrt{-(p_1-p_2)^2}$
 and
 $\tilde{f}(Q)$ is the Fourier transform (characteristic function) of $f(x)$.
The parameter $\gamma$ and
the $(1 + \delta Q)$ term have been
introduced to parametrize
possible long-range correlations not adequately accounted for in the reference sample,
and the parameter $\lambda$ to account for several factors, such as the possible lack of complete
incoherence of particle production and the presence of long-lived resonance decays if the particle emission
consists of a small, resolvable core and a halo with experimentally unresolvable large length scales
\cite{Bolz:1992hc,Csorgo:1994in}.

%\begin{equation}
%  \tilde{f} (Q) = \int \mathrm{d} x \exp (iQx) f(x).
%\end{equation}
 
\subsection{Gaussian distributed source}            \label{sect:Gauss}
The simplest assumption is that the source has a symmetric Gaussian distribution,
in which case
$\tilde{f} (Q)=\exp \left(i\mu Q -\frac{(RQ)^2}{2}\right)$ and
\begin{equation} \label{eq:gaussR2}
  R_2(Q) = \gamma \left[1 + \lambda \exp \left(-(R Q)^2 \right) \right]
           \left(1 + \delta Q \right) \;.
\end{equation}
 
\begin{figure}
  \centering
  \includegraphics[width=.75\figwidth]{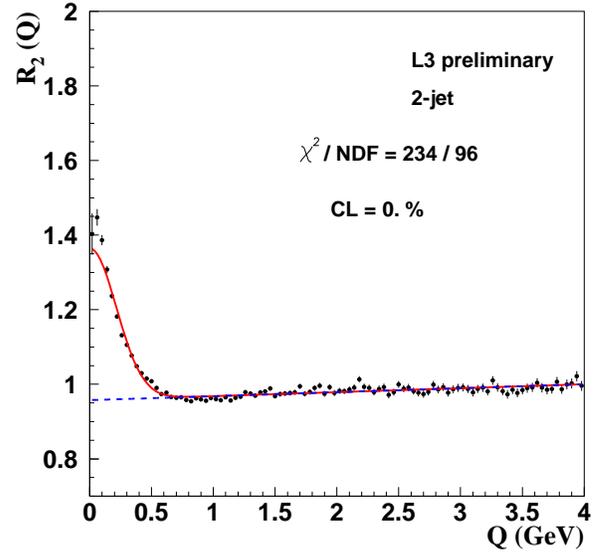}
  \includegraphics[width=.75\figwidth]{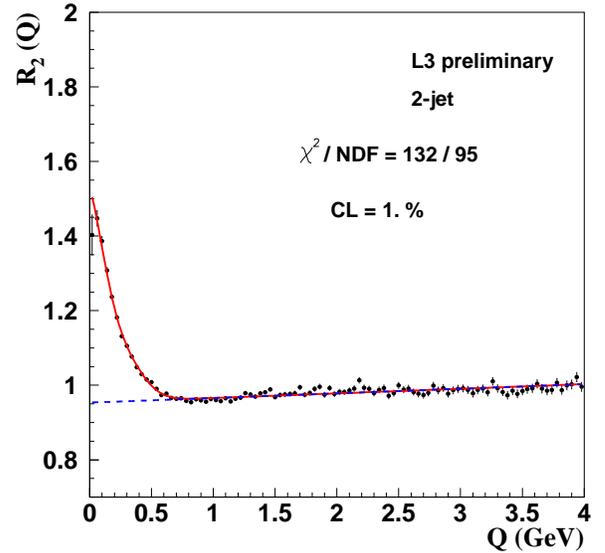}
  \caption{The Bose-Einstein correlation function $R_2$ for two-jet events with the result of a fit of
           (a) the Gaussian and (b) the Edgeworth parametrizations,
           \Eqs{eq:gaussR2} and (\ref{eq:edgeworthR2}), respectively.
           The dashed line represents the long-range part of the fit, \ie, $\gamma(1+\delta Q)$.
           \label{fig:gauss_2jet}
           }
\end{figure}

\begin{figure}
  \centering
  \includegraphics[width=.75\figwidth]{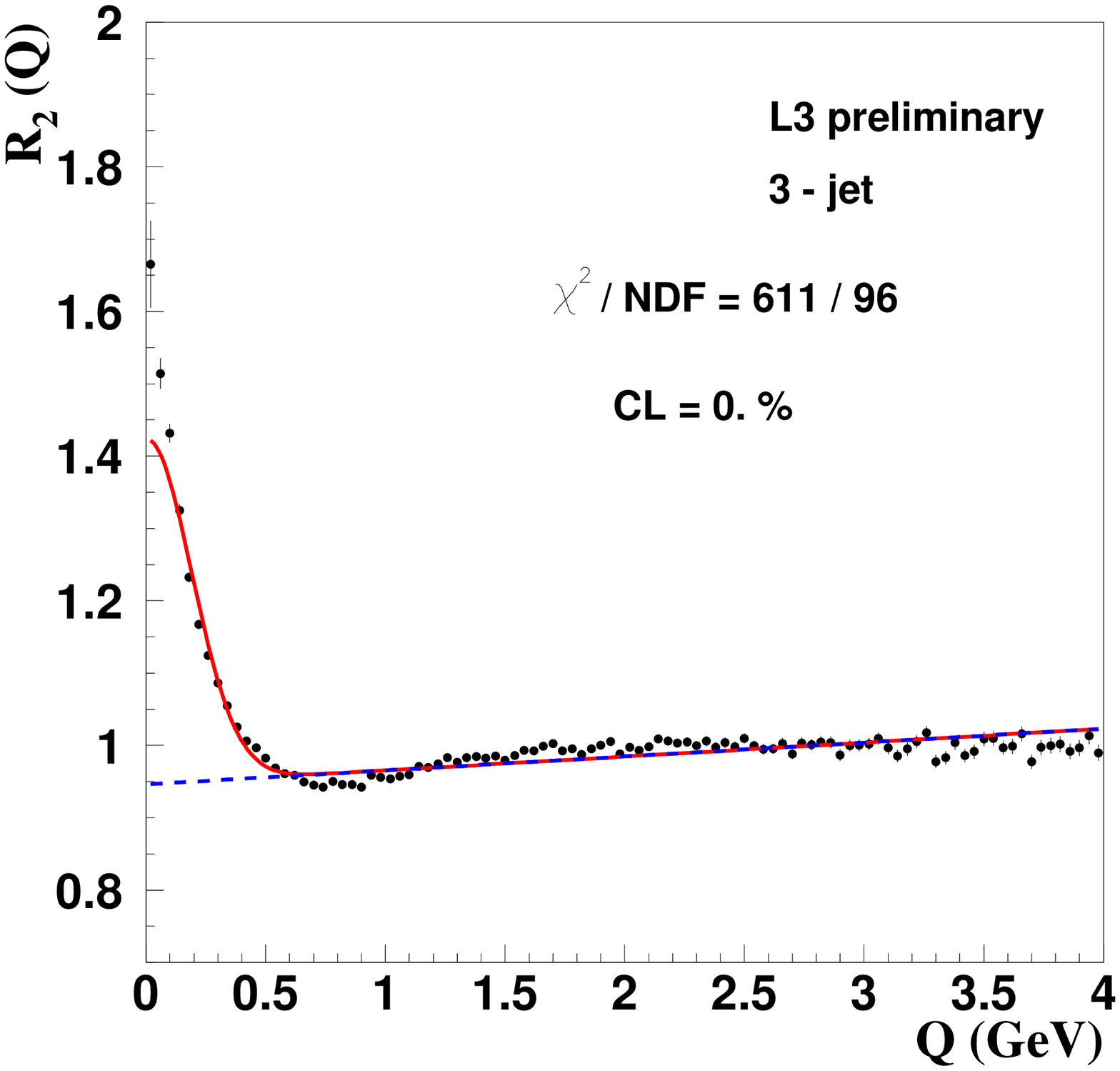}
  \includegraphics[width=.75\figwidth]{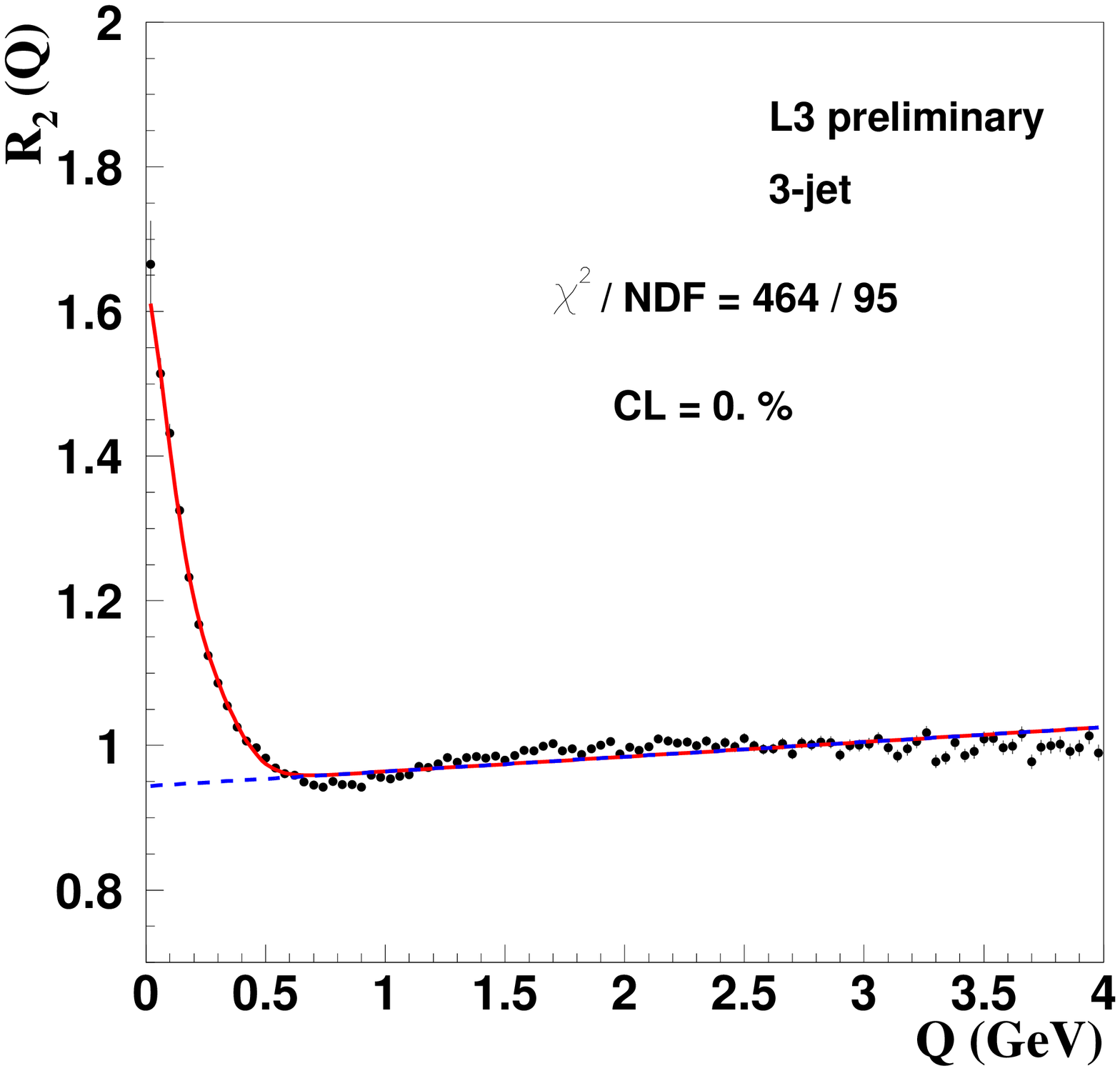}
  \caption{The Bose-Einstein correlation function $R_2$ for three-jet events with the result of a fit of
            (a) the Gaussian and (b) the Edgeworth parametrizations,
            \Eqs{eq:gaussR2} and (\ref{eq:edgeworthR2}), respectively.
           The dashed line represents the long-range part of the fit, \ie, $\gamma(1+\delta Q)$.
           \label{fig:gauss_3jet}
           }
\end{figure}

A fit of \Eq{eq:gaussR2} to the data results in an unacceptably low confidence level.
The fit is particularly bad at low $Q$ values, as is shown in \Fig{fig:gauss_2jet}a for two-jet events
and in \Fig{fig:gauss_3jet}a for three-jet events,
%and in \Fig{fig:gauss_all}a for all events,
from which we conclude that the shape of the source deviates from a Gaussian.

A model-independent way to study deviations from the Gaussian parametrization is to
use~\cite{Tamas:Moriond28,Tamas:Cracow1994,Tamas:HIP2002}
the Edgeworth expansion~\cite{Edgeworth} about a Gaussian. %, for which a Gaussian is the lowest-order term.
Keeping only the first non-Gaussian term, we have
\begin{equation} \label{eq:edgeworthR2}
  R_2(Q) = \gamma \left(1 + \lambda \exp \left(-(R Q)^2 \right)
                  \left[1+\frac{\kappa}{3!}H_{3}(RQ)\right]
                  \right)
           \left(1 + \delta Q \right) \;,
\end{equation}
where
$\kappa$ is the third-order cumulant moment and
$H_{3}(RQ)\equiv (\sqrt{2}RQ)^{3}-3\sqrt{2}RQ$
is the third-order Hermite polynomial.
Note that the second-order cumulant corresponds to the radius~$R$.
 
A fit of \Eq{eq:edgeworthR2} to the two-jet data, shown in \Fig{fig:gauss_2jet}b, is indeed
much better than the purely Gaussian fit.  However, the confidence level is still marginal, and close
inspection of the figure shows that the fit curve is systematically above the data in the region
0.6--1.2\,\GeV\ and that the data for $Q\ge1.5$\,\GeV\ appear flatter than the curve, as is also the case
for the purely Gaussian fit.
Similar behavior is observed for three-jet events
(\Fig{fig:gauss_3jet}b)
and for all events.
%(\Fig{fig:gauss_all}b).

\subsection{L\'evy distributed source}         \label{sect:Levy}
%Adopting Nolan's $S(\alpha, \beta=0, \gamma, \delta; 1)$ convention~\cite{Nolan} for the symmetric
%L\'evy stable distribution with rescaling of the scale parameter $\gamma$ to $R$ and
%the location parameter $\delta$ to $x_0$, the Fourier transform (characteristic function)
The symmetric L\'evy stable distribution is characterized by three parameters: $x_0$, $R$, and $\alpha$.
Its Fourier transform,
$\tilde{f}(Q)$, has the following form:
\begin{equation}
% \tilde{f}(Q) = \exp (iQ x_0 - |R Q|^\alpha) \;.
  \tilde{f}(Q) = \exp \left(iQ x_0 - \frac{|R Q|^\alpha}{2} \right) \;.
\end{equation}
The index of stability, $\alpha$, satisfies the inequality $0<\alpha \leq 2$.
The case $\alpha=2$ corresponds to a Gaussian source distribution with mean $x_0$ and standard deviation $R$.
For more details, see, \eg, \cite{Nolan}.
 
\begin{figure}
  \centering
  \includegraphics[width=.75\figwidth]{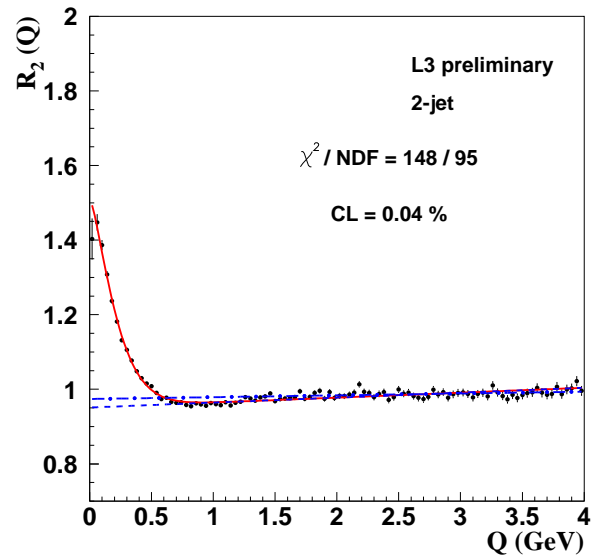}
  \caption{The Bose-Einstein correlation function $R_2$ for two-jet events. The curve
           corresponds to the fit of the symmetric L\'evy parametrization, \Eq{eq:symlevR2}.
           The dashed line represents the long-range part of the fit, \ie, $\gamma(1+\delta Q)$.
           The dot-dashed line represents a linear fit in the region $Q>1.5$\,\GeV.
           \label{fig:sym_levy_2jet}
           }
\end{figure}

\begin{figure}
  \centering
  \includegraphics[width=.75\figwidth]{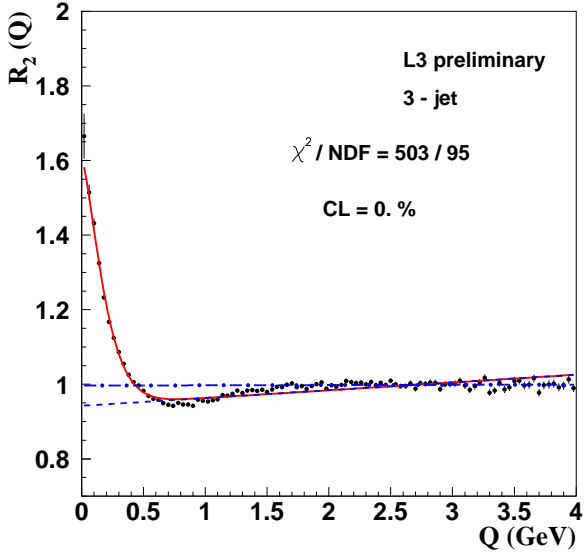}
  \caption{The Bose-Einstein correlation function $R_2$ for three-jet events. The curve
           corresponds to the fit of the symmetric L\'evy parametrization, \Eq{eq:symlevR2}.
           The dashed line represents the long-range part of the fit, \ie, $\gamma(1+\delta Q)$.
           The dot-dashed line represents a linear fit in the region $Q>1.5$\,\GeV.
           \label{fig:sym_levy_3jet}
           }
\end{figure}

Then $R_2$ has the following, relatively simple, form~\cite{Tamas:Levy2004}:
\begin{equation}\label{eq:symlevR2}
    R_2(Q) = \gamma \left[ 1+ \lambda \exp \left(-(RQ)^\alpha \right) \right]
             (1+ \delta Q) \;.
\end{equation}
From the fit of \Eq{eq:symlevR2} to the two-jet data, shown in \Fig{fig:sym_levy_2jet}, it is clear
that the correlation function is far from Gaussian: $\alpha = 1.34\pm0.04$.
The confidence level, although
improved compared to the fit of \Eq{eq:gaussR2}, is still unacceptably low,
in fact worse than that for the Edgeworth parametrization.
The same is true for three-jet events
(\Fig{fig:sym_levy_3jet})
and for all events.
%(\Fig{fig:sym_levy_all}).
The values of $\alpha$ are $1.39\pm0.04$ for three-jet and $1.43\pm0.03$ for all events, respectively.
%\mnote{?}\enote{I don't think we need these values in the actual paper.}

Both the symmetric L\'evy parametrization and the Edgeworth parametrizations do a fair job of describing the
region $Q<0.6\,\GeV$, but fail at higher $Q$.  $R_2$ in the region $Q\ge1.5\,\GeV$ is nearly constant
($\approx1$). However, in the region 0.6--1.5\,\GeV\ $R_2$ has a smaller value,
dipping below unity~\footnote{More correctly, dipping below the value of the parameter $\gamma$.}, which is
indicative of an anti-correlation.
This is clearly seen
in \Figs{fig:sym_levy_2jet} and~\ref{fig:sym_levy_3jet}
by comparing the data in this region  to an extrapolation of a linear fit, \Eq{eq:symlevR2} with
$\lambda=0$, in the region $Q\ge1.5\,\GeV$.
%which is indicated by the dashed line
The inability to describe this dip in $R_2$ is the primary reason for the failure of both the
Edgeworth and symmetric L\'evy parametrizations.

\subsection{Time dependence of the source}
The parametrizations discussed so far, which have proved insufficient to describe the BEC, all assume a
static source.  The parameter $R$, representing the size of the source as seen in the rest frame of the
pion pair, is a constant.  It has, however, been observed that $R$ depends on the transverse mass,
$\mt=\sqrt{m^2+\pt^2}=\sqrt{E^2-p_z^2}$, of the pions~\cite{Smirnova:Nijm96,Dalen:Maha98}.
It has been shown~\cite{Bialas:1999,Bialas:2000} that this dependence can be understood if the produced
pions satisfy, approximately, the (generalized) Bjorken-Gottfried
condition~\cite{Gottfried:1972,Bjorken:SLAC73,Bjorken:1973,Gottfried:1974,Low:1978,Bjorken:ISMD94}, whereby
the four-momentum of a produced particle and the space-time position at which it is produced are linearly
related:
\begin{equation} \label{eq:GB-corr}
   x^\mu = d k^\mu \;.
\end{equation}
Such a correlation between space-time and momentum-energy is  also a feature of the Lund string model
as incorporated in \JETSET,
which is very successful in describing detailed features of the hadronic final states of \Pep\Pem\ annihilation.
 
In the previous section we have seen that BEC depend, at least approximately, only on $Q$ and not on
its components separately.
This is a non-trivial result.
For a hydrodynamical type of source,
on the contrary, BEC decrease when any of the relative momentum components is large~\cite{Tamas:HIP2002, Tamas;Lorstad:1996}.
Further, we have seen that $R_2$ in the region 0.6--1.5\,\GeV\ dips below its values at higher $Q$.
 
A model which predicts such $Q$-dependence while incorporating the  Bjorken-Gottfried condition
is the so-called $\tau$-model, described below.
 
\subsubsection{The \boldmath{$\tau$}-model}  \label{sect:taumodel}
A model of strongly correlated phase-space, known as the $\tau$-model~\cite{Tamas;Zimanji:1990},
explains the experimentally found invariant relative momentum
dependence of Bose-Einstein correlations in $\mathrm{e}^+\mathrm{e}^-$ reactions.
This model also predicts a specific transverse mass dependence of $R_2$,
that we subject to an experimental test here.
 
In this model, it is assumed that the average production point in the overall center-of-mass system,
$\overline{x}=(\overline{t},\overline{r}_x,\overline{r}_y,\overline{r}_z)$, of particles with a given
four-momentum $k$ is given by
\begin{equation} \label{eq:tau-corr}
   \overline{x}^\mu (k^\mu)  = d k^\mu \;.
\end{equation}
In the case of two-jet events,
\begin{equation} \label{eq:d}
   d=\tau/\mt \;,  %   d = \frac{\tau}{\mt} \;,
\end{equation}
where
% $\mt=\sqrt{m^2+\pt^2}=\sqrt{E^2-p_z^2}$ is the transverse mass
\mt\ is the transverse mass
and
$\tau = \sqrt{\overline{t}^2 - \overline{r}_z^2}$ is the longitudinal proper time~\footnote{The
terminology `longitudinal' proper time and `transverse' mass seems customary in the literature
even though their definitions are analogous $\tau = \sqrt{\overline{t}^2 - \overline{r}_z^2}$
and                                         $ \mt = \sqrt{E^2            - p_z^2}$.}.
For isotropically distributed particle production, the transverse mass is replaced by the
mass in \Eq{eq:d}, while for the case of three-jet events the relation is more complicated.
%\mnote{!}\enote{Perhaps phrase this differently.}
The second assumption is that the distribution of $x^\mu (k^\mu)$ about its average,
$\delta_\Delta ( x^\mu(k^\mu) - \overline{x}^\mu (k^\mu) )$, is narrower than the
proper-time distribution.
Then the emission function of the $\tau$-model is
\begin{equation} \label{eq:source}
  S(x,k) = \int_0^{\infty} \mathrm{d}\tau H(\tau)\delta_{\Delta}(x-dk) \rho_1(k) \;,
\end{equation}
where $H(\tau)$ is the longitudinal proper-time distribution, the factor
$\delta_{\Delta}(x-dk)$ describes the strength of the correlations between
coordinate space and momentum space variables and $\rho_1(k)$ is the experimentally
measurable single-particle spectrum.
 
The two-pion distribution, $\rho_2(k_1,k_2)$, is related to $S(x,k)$, in the plane-wave approximation,
by the Yano-Koonin formula~\cite{Yano}:
%In the plane-wave approximation, the Yano-Koonin formula~\cite{Yano} gives the following two-pion distribution:
\begin{eqnarray}  \label{eq:yano}
   \rho_2(k_1,k_2) &=& \int \mathrm{d}^4 x_1 \mathrm{d}^4 x_2 S(x_1,k_1) S(x_2,k_2)  \nonumber \\
& & \cdot             \left(\strut 1 + \cos\left(\strut\left[k_1-k_2\right]\left[x_1-x_2\right]\right)\strut\right)\;.
\end{eqnarray}
Approximating the function $\delta_\Delta$ by a Dirac delta function,
the argument of the cosine becomes
\begin{equation}
   (k_1 - k_2)(\bar{x}_1 - \bar{x}_2) = - 0.5 (d_1 + d_2) Q^2 \;.
\end{equation}
Then the two-particle Bose-Einstein correlation function is approximated by
\begin{equation}     \label{eq:levy2jetR2a}
   R_2(k_1,k_2) = 1 + \lambda \mathrm{Re} \widetilde{H}^2
                  \left(  \frac{Q^2}{2 \overline{m}_{\mathrm{t}}} \right) \;,
\end{equation}
where $\widetilde{H}(\omega) = \int \mathrm{d} \tau H(\tau) \exp(i \omega \tau)$
is the Fourier transform of $H(\tau)$.
Thus an invariant relative momentum dependent BEC appears.
Note that $R_2$ depends not only on $Q$ but also on the average transverse mass of the two pions,
$\overline{m}_\mathrm{t}$.

Since there is no particle production before the onset of the collision,
$H(\tau)$ should be a  one-sided distribution.
We choose a one-sided L\'evy distribution, which has the characteristic function~\cite{Tamas:Levy2004}
(for $\alpha\ne1$)
\begin{equation} \label{eq:levy1sidecharf}
   \widetilde{H}(\omega) = \exp\left[ -\frac{1}{2}\left(\Delta\tau|\omega|\strut\right)^\alpha
          \left( 1 -  i\, \mathrm{sign}(\omega)\tan\left(\frac{\alpha\pi}{2}\right) \strut \right)
       + i\,\omega\tau_0\right]
%                                  \;,\quad \alpha\ne1   \;,
\end{equation}
where the parameter $\tau_0$ is the proper time of the onset of particle production
and $\Delta \tau$ is a measure of the width of the proper-time distribution.
For the special case $\alpha=1$, see, \eg, \cite{Nolan}.
Using this characteristic function in \Eq{eq:levy2jetR2a} yields
\begin{eqnarray} \label{eq:levy2jetR2}
   R_2(Q,\overline{m}_\mathrm{t}) &=&
       \gamma
       \left[ 1+\lambda \cos \left( \frac{\tau_0 Q^2}{\overline{m}_\mathrm{t}}
                                   + \tan \left( \frac{\alpha \pi}{2} \right)
                                     \left( \frac{\Delta\tau Q^2}{2\overline{m}_\mathrm{t}} \right)^{\!\alpha} \right)
   \right. \nonumber \\
   & & \cdot\left.      \exp \left( -\left( \frac{\Delta\tau Q^2}{2\overline{m}_\mathrm{t}} \right)^{\!\alpha} \right)
        \right] (1+\delta Q)   \;.
\end{eqnarray}

\subsubsection{The \boldmath{$\tau$}-model for average \boldmath{$\mt$}}  \label{sect:tausimpl}
\begin{figure}
  \centering
  \includegraphics[width=.75\figwidth]{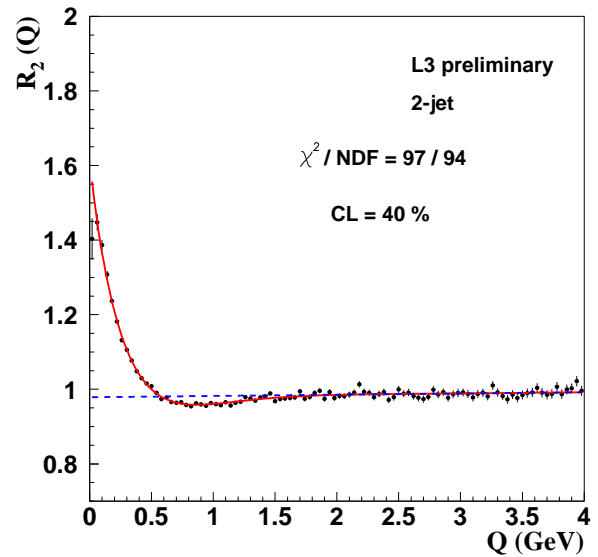}
  \caption{The Bose-Einstein correlation function $R_2$ for two-jet events. The curve
           corresponds to the fit of the one-sided L\'evy parametrization, \Eq{eq:asymlevR2}.
           The dashed line represents the long-range part of the fit, \ie, $\gamma(1+\delta Q)$.
           \label{fig:a_levy_2jet}
           }
\end{figure}

\begin{figure}
  \centering
  \includegraphics[width=.75\figwidth]{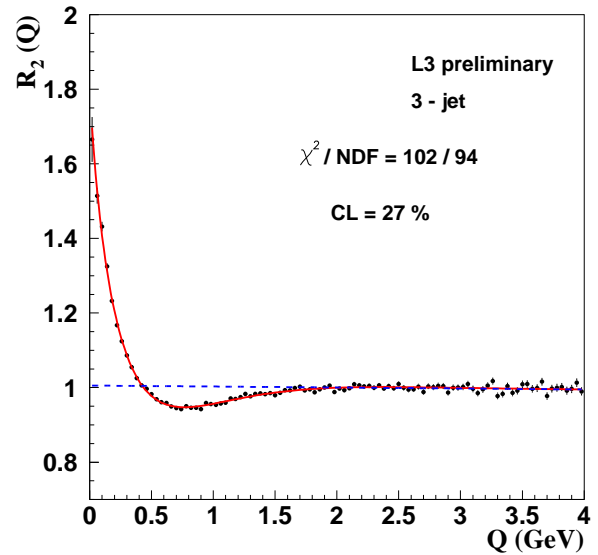}
  \caption{The Bose-Einstein correlation function $R_2$ for three-jet events. The curve
           corresponds to the fit of the one-sided L\'evy parametrization, \Eq{eq:asymlevR2}.
           \label{fig:a_levy_3jet}
           The dashed line represents the long-range part of the fit, \ie, $\gamma(1+\delta Q)$.
           }
\end{figure}

Before proceeding to fits of \Eq{eq:levy2jetR2}, we first consider a simplification of the equation
obtained by assuming (a) that particle production starts immediately, \ie, $\tau_0=0$,
and (b) an average \mt-dependence, which is implemented in an approximate way by defining an effective
radius, $R=\sqrt{\Delta\tau/(2\overline{m}_\mathrm{t})}$.
This results in:
\begin{equation}\label{eq:asymlevR2}
    R_2(Q) = \gamma \left[ 1+ \lambda \cos \left[(R_\mathrm{a}Q)^{2\alpha} \right]
             \exp \left(-(RQ)^{2\alpha} \right) \right] (1+ \delta Q) \;,
\end{equation}
where $R_\mathrm{a}$ is  related to $R$ by
\begin{equation}\label{eq:asymlevRaR}
    R_\mathrm{a}^{2\alpha} = \tan\left(\frac{\alpha\pi}{2}\right) R^{2\alpha} \;.
\end{equation}
Fits of
\Eq{eq:asymlevR2} are first performed with $R_\mathrm{a}$ as a free parameter.
The fit results obtained, for two-jet, three-jet, and all events are listed in \Tab{tab:a_levy}
and shown  in \Fig{fig:a_levy_2jet} for two-jet events and in \Fig{fig:a_levy_3jet} for three-jet events.
%, \ref{fig:a_levy_3jet}, and \ref{fig:a_levy_all}, respectively.
They have acceptable confidence levels, describing well the dip below unity in the 0.6--1.5\,\GeV\ region,
as well as the low-$Q$ peak.

The fit parameters for the two-jet events satisfy \Eq{eq:asymlevRaR}.
However, those for three-jet and all events do not.
We note that the values of the parameters $\alpha$ and $R$  % , and $R_\mathrm{a}$
do not differ greatly between
2- and 3-jet samples, the most significant difference appearing to be nearly 3$\sigma$ for $\alpha$.
However, these parameters are rather highly correlated
(in the fit for all events, the correlation coefficients are $\rho(\lambda,R)=0.95$,
$\rho(\lambda,\alpha)=-0.67$ and $\rho(R,\alpha)=-0.61$,
which makes the simple calculation of the
statistical significance of differences in the parameters unreliable.

Fit results imposing \Eq{eq:asymlevRaR} are given in \Tab{tab:a_levy_c}.
%and shown in \Figs{fig:a_levy_2jet_c}, \ref{fig:a_levy_3jet_c}, and \ref{fig:a_levy_all_c}.
For two-jet events, the values of the parameters are the same as in the fit with $R_\mathrm{a}$ free---only
the uncertainties have changed.
For three-jet and all events, the imposition of \Eq{eq:asymlevRaR} results in values of $\alpha$ and $R$
closer to those for two-jet events, but the confidence levels are very bad, a consequence of
incompatibility with \Eq{eq:asymlevRaR},
an incompatibility that is not surprizing given that \Eq{eq:d} is only valid for two-jet events.
Therefore, we only consider two-jet events in the remaining sections of this \Letter.

\begin{table}[htbp]
\caption{Results of fits of \Eq{eq:asymlevR2}
           for two-jet, three-jet, and all events.
  The uncertainties are only statistical.
%\enote{Check values.}
         }
\centering
\begin{tabular}{ l r@{$\;\pm\;$}l
                   r@{$\;\pm\;$}l
                   r@{$\;\pm\;$}l
               }
\hline
    parameter        & \multicolumn{2}{c }{2-jet} & \multicolumn{2}{c }{3-jet} & \multicolumn{2}{c }{ all}
\\ \hline
    $\alpha$         & 0.42  & 0.02               &  0.35  & 0.01               & 0.38  & 0.01       \\
    $\lambda$        & 0.67  & 0.03               &  0.84  & 0.04               & 0.73  & 0.02       \\
 $R$            (fm) & 0.79  & 0.04               &  0.89  & 0.03               & 0.81  & 0.03       \\
 $R_\mathrm{a}$ (fm) & 0.59  & 0.03               &  0.88  & 0.04               & 0.81  & 0.02       \\
    $\delta$         & 0.003 & 0.002              &--0.003 & 0.002              & 0.003 & 0.001      \\
    $\gamma$         & 0.979 & 0.005              &  1.001 & 0.005              & 0.997 & 0.003      \\
    \hline
  \chisq/DoF         & \multicolumn{2}{c }{97/94} & \multicolumn{2}{c }{102/94}& \multicolumn{2}{c }{98/94}
\\
  confidence level   & \multicolumn{2}{c }{40\%}  & \multicolumn{2}{c }{27\%}  & \multicolumn{2}{c }{37\%}
\\ \hline
\end{tabular}
\label{tab:a_levy}
\end{table}
 
\begin{table}[htbp]
\caption{Results of fits of \Eq{eq:asymlevR2} imposing \Eq{eq:asymlevRaR}
           for two-jet, three-jet, and all events.
  The uncertainties are only statistical.
%\enote{Check values.}
         }
\centering
\begin{tabular}{ l r@{$\;\pm\;$}l
                   r@{$\;\pm\;$}l
                   r@{$\;\pm\;$}l
               }
\hline
    parameter        & \multicolumn{2}{c }{2-jet} & \multicolumn{2}{c }{3-jet} & \multicolumn{2}{c }{ all} \\
\hline
    $\alpha$         & 0.42  & 0.01               &  0.44  & 0.01               & 0.45  & 0.01       \\
    $\lambda$        & 0.67  & 0.03               &  0.77  & 0.04               & 0.69  & 0.03       \\
 $R$            (fm) & 0.79  & 0.03               &  0.84  & 0.04               & 0.79  & 0.03       \\
    $\delta$         & 0.003 & 0.001              &  0.010 & 0.001              & 0.009 & 0.001      \\
    $\gamma$         & 0.979 & 0.005              &  0.972 & 0.001              & 0.973 & 0.001      \\
\hline
  \chisq/DoF         & \multicolumn{2}{c }{97/95} & \multicolumn{2}{c }{174/95}& \multicolumn{2}{c }{175/95} \\
  confidence level   & \multicolumn{2}{c }{42\%} & \multicolumn{2}{c }{$10^{-6}$} & \multicolumn{2}{c }{$10^{-6}$} \\
\hline
\end{tabular}
\label{tab:a_levy_c}
\end{table}
 
%\enote{Need systematics.}\mnote{syst}

\subsubsection{The \boldmath{$\tau$}-model with \boldmath{$\mt$} dependence}  \label{sect:taufits}
 
\begin{figure}
  \centering
  \includegraphics*[width=.80\figwidth]{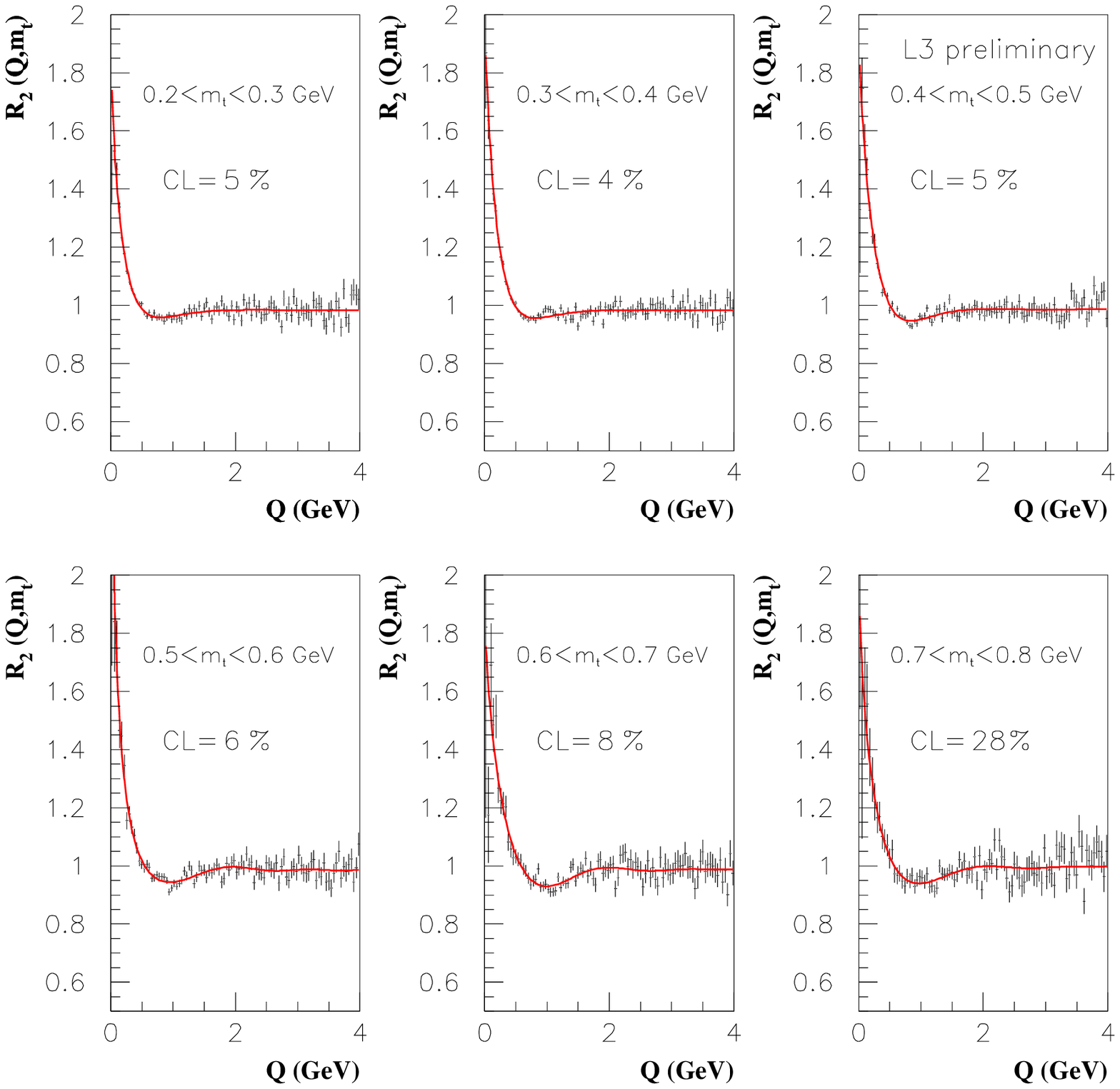}
  \caption{The results of fits of \Eq{eq:levy2jetR2} to two-jet data for various intervals of \mt.
  \label{fig:2jetR2}
  }
\end{figure}

\begin{figure}%[htbp]
  \centering
  \includegraphics*[width=.75\figwidth]{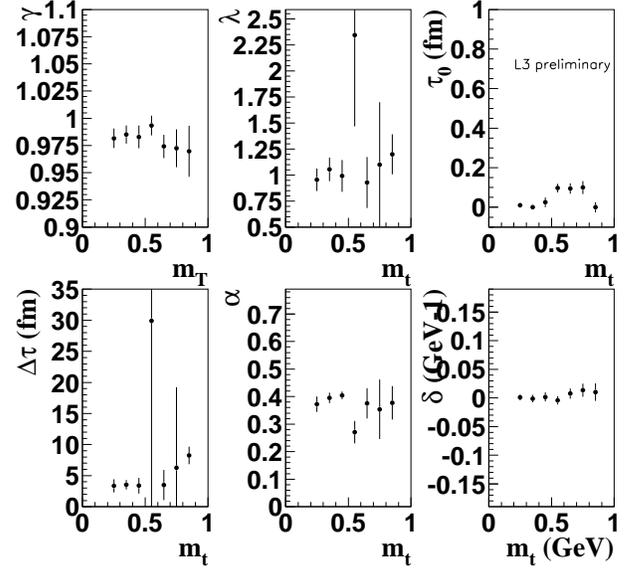}
  \caption{The fit parameters from fits of \Eq{eq:levy2jetR2} to two-jet data for various intervals of \mt.
  \label{fig:mtfitpars}
  }
\end{figure}

Fits of \Eq{eq:levy2jetR2} to the two-jet data are performed in several \mt\ intervals.
The  resulting fits are shown for several \mt\ intervals in \Fig{fig:2jetR2}, and
the values of the parameters obtained in the fits
are listed in \Fig{fig:mtfitpars}.
The quality of the fits
is seen to be statistically acceptable and the fitted values of the model parameters,
$\alpha$, $\tau_0$ and $\Delta\tau$,
are stable
and within errors independent of \mt, confirming the expectation of the $\tau$-model.
%\mnote{??}\enote{Well, with the exception of one crazy point.}
We conclude that the $\tau$-model with a one-sided Levy proper-time distribution describes the data
with parameters $\tau_0\approx 0$ fm, $\alpha \approx 0.38\pm 0.05$ and $\Delta \tau \approx 3.5\pm0.6$ fm.
These values are consistent with the fit of \Eq{eq:asymlevR2} in the previous section, including the value
of $R$, which, combined with the average value of \mt\ (0.563\,\GeV), corresponds to $\Delta\tau=3.5$\,fm.
Just as in the fit of \Eq{eq:asymlevR2}, the parameters of the L\'evy distribution are highly correlated.
Typical values
of the correlation coefficients are $\rho(\lambda,\Delta\tau)=0.95$,
$\rho(\lambda,\alpha)=-0.67$ and $\rho(\Delta\tau,\alpha)=-0.9$.

\section{The emission function of two-jet events}  \label{sect:emission2jet}
Within the framework of the $\tau$-model, we now
reconstruct the space-time picture of the emitting process for two-jet events.
The emission function in configuration space, $S(x)$, is the proper time derivative of the
integral over $k$ of $S(x,k)$, which in the
$\tau$-model is given by \Eq{eq:source}.
Approximating $\delta_\Delta$ by a Dirac delta function, we find
\begin{equation}   \label{eq:Sspace}
   S(x) = \frac{\mathrm{d}^4 n}{\mathrm{d}\tau\mathrm{d}^3 r}
        = \left(\frac{\mt}{\tau}\right)^3 H(\tau) \rho_1\left( k=\frac{\mt r}{\tau} \right) \;.
\end{equation}
 
To simplify the reconstruction of $S(x)$ we assume that
%the emission function in configuration space
it
can be factorized in the following way:
\begin{equation}   \label{eq:fact}
     S(r,z,t) = I(r) G(\eta) H(\tau) \;,
\end{equation}
where $I(r)$ is the single-particle  transverse distribution, $G(\eta)$ is
the space-time rapidity distribution of particle production,
and $H(\tau)$ is the proper-time distribution.
With the strongly correlated phase-space of the $\tau$-model,
$\eta=y$ and $r=\pt\tau/\mt$.
%$G(\eta)$ approximately coincides with the single-particle rapidity distribution,
%and $I(r)$ is closely related to the \pt\ distribution:
Hence,
\begin{eqnarray}
      G(\eta)   &=& N_y(\eta)                                           \;, \label{eq:Geta} \\
      I(r     ) &=& \left(\frac{\mt}{\tau}\right)^{\!3} N_{\pt}(r\mt/ \tau) \;, \label{eq:Ir}
\end{eqnarray}
where $N_y$ and $N_{\pt}$ are the single-particle inclusive rapidity and \pt\ distributions, respectively.
%The proper-time distribution, $H(\tau)$, is found from the fit of \Eq{eq:levy2jetR2} in previous section.
The factorization of transverse and longitudinal distributions has been checked.
The distribution of \pt\ is, to a good approximation,
independent of the rapidity \cite{tamas}.
%This is shown in \Fig{fig:fac}.
 
\begin{figure}%[thbp]
  \centering
   \includegraphics*[width=.55\figwidth]{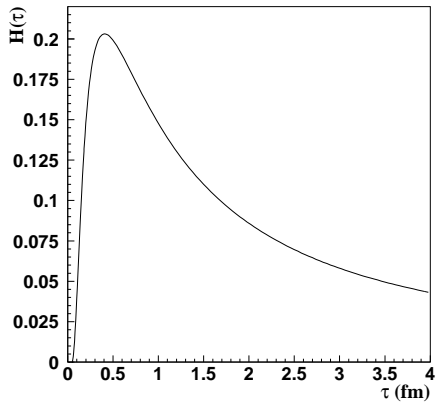}
  \caption{The proper time distribution, $H(\tau)$,
           for $\alpha=0.4$, $\tau_0=0$ and $\Delta\tau=3.5$\,fm.
  \label{fig:Htau}
  }
\end{figure}
 
\begin{figure}
  \centering
  \includegraphics*[width=.75\figwidth]{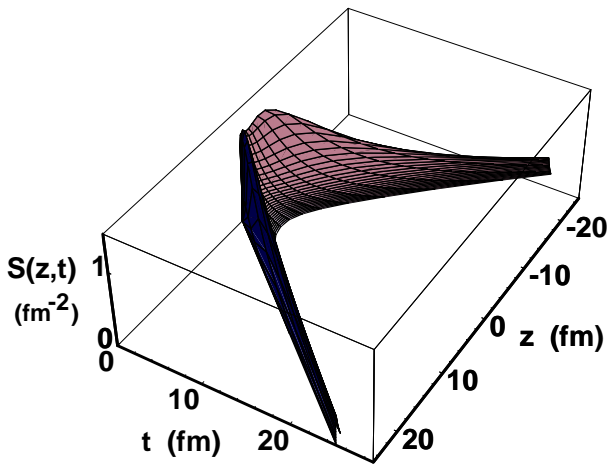} % \\ % \hfil
  \caption{The temporal-longitudinal part of the source function normalized to the
           average number of pions per event.
\label{fig:longemis}
          }
\end{figure}

With these assumptions and using $H(\tau)$ as obtained from the fit of \Eq{eq:levy2jetR2}
(shown in \Fig{fig:Htau})
together with the inclusive rapidity and \pt\ distributions \cite{tamas},
%\enote{(shown in \Fig{fig:yptincl}---not to be included in the paper, but is here for possible discussion)}
the full emission function is reconstructed.
Its integral over the transverse distribution is plotted in \Fig{fig:longemis}.
It exhibits a ``boomerang'' shape with a maximum at low $t$ and $z$
but with tails reaching out to very large values of $t$ and $z$,
a feature also observed in hadron-hadron~\cite{NA22emiss} and heavy ion collisions~\cite{HIemiss:Ster}.
 
\begin{figure}
  \centering
  \includegraphics[width=.39\figwidth]{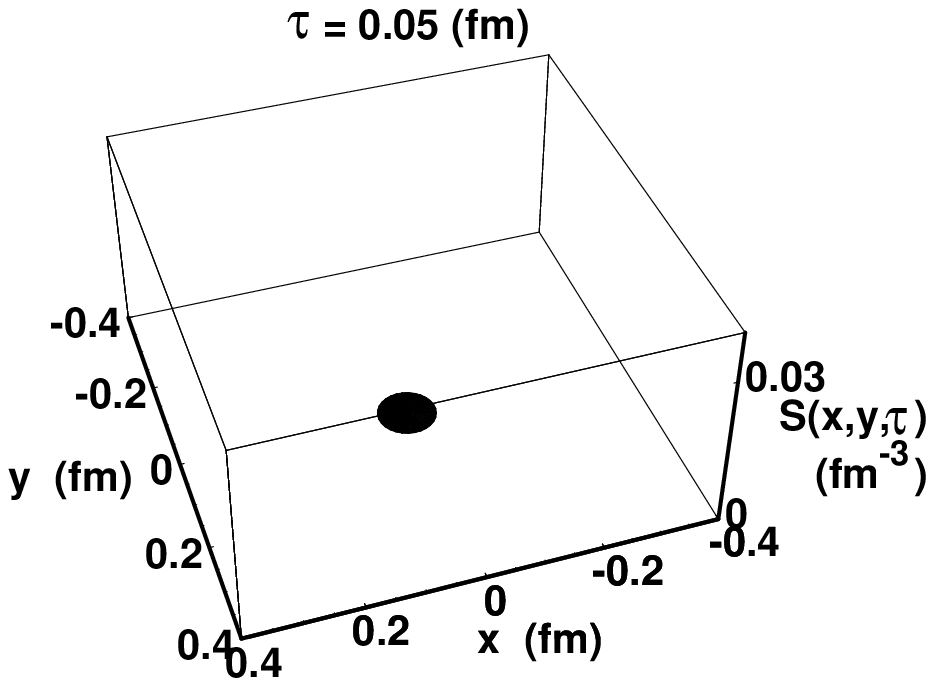}
  \includegraphics[width=.39\figwidth]{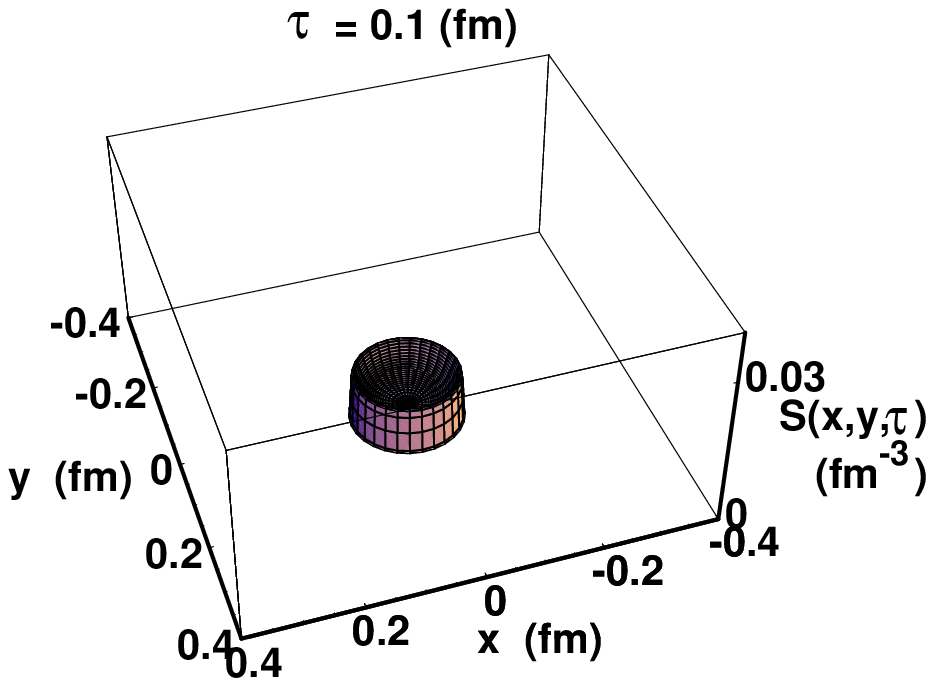}
  \includegraphics[width=.39\figwidth]{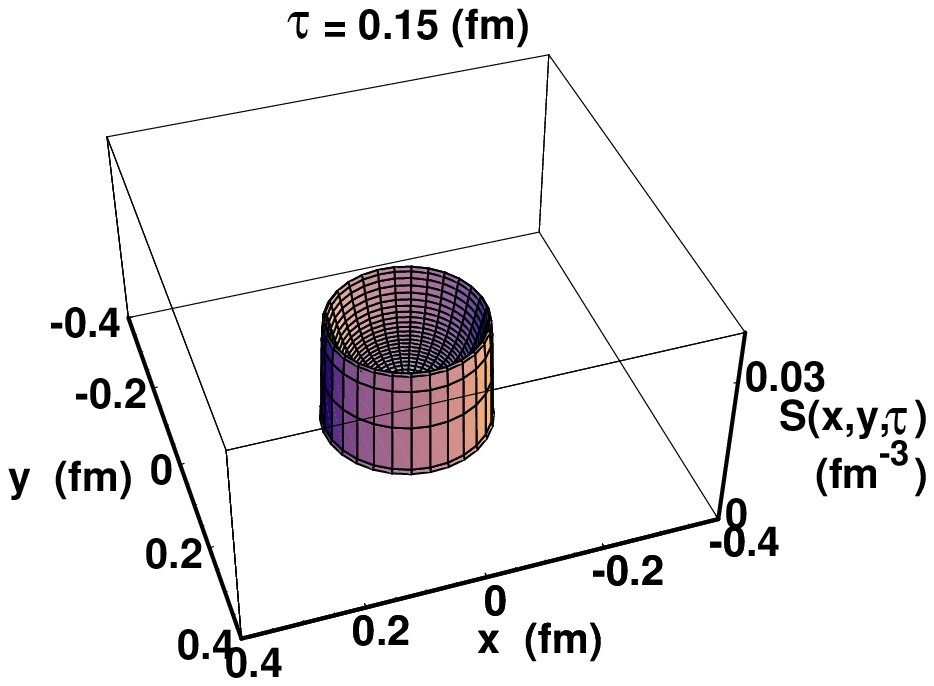}
  \includegraphics[width=.39\figwidth]{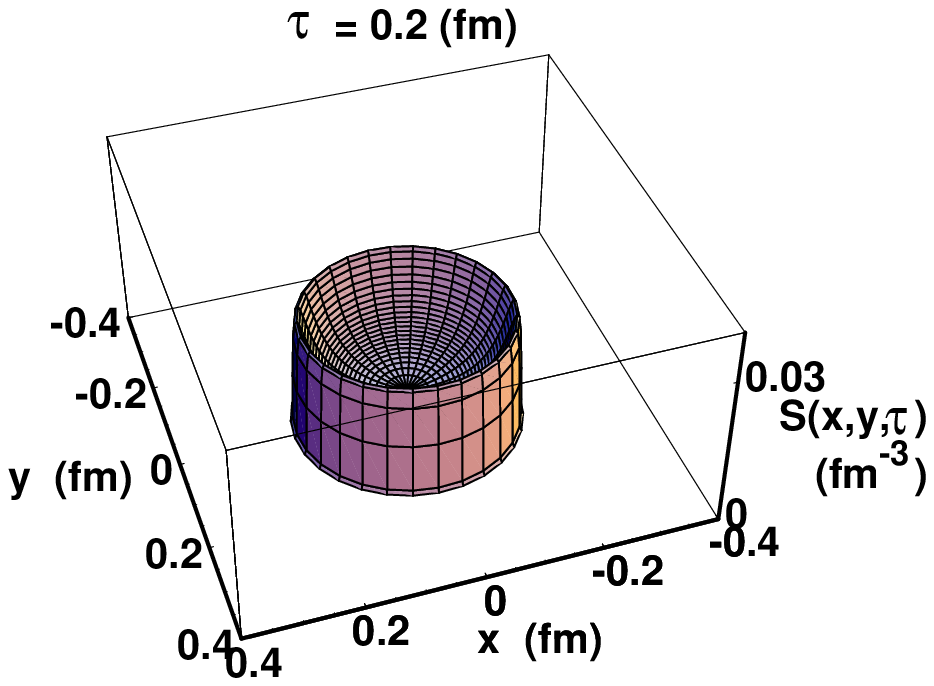}
  \includegraphics[width=.39\figwidth]{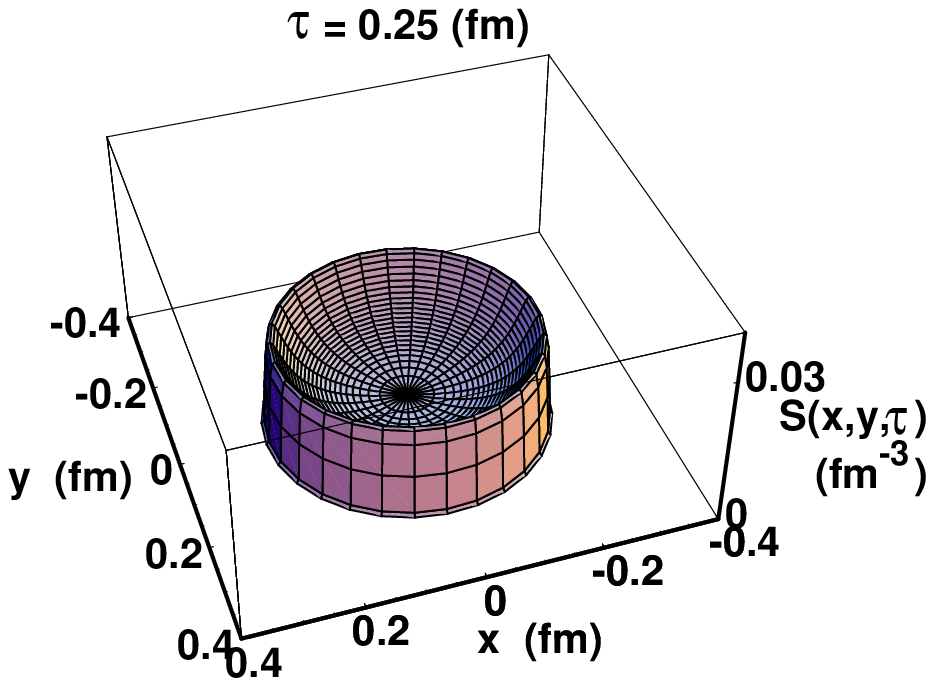}
  \includegraphics[width=.39\figwidth]{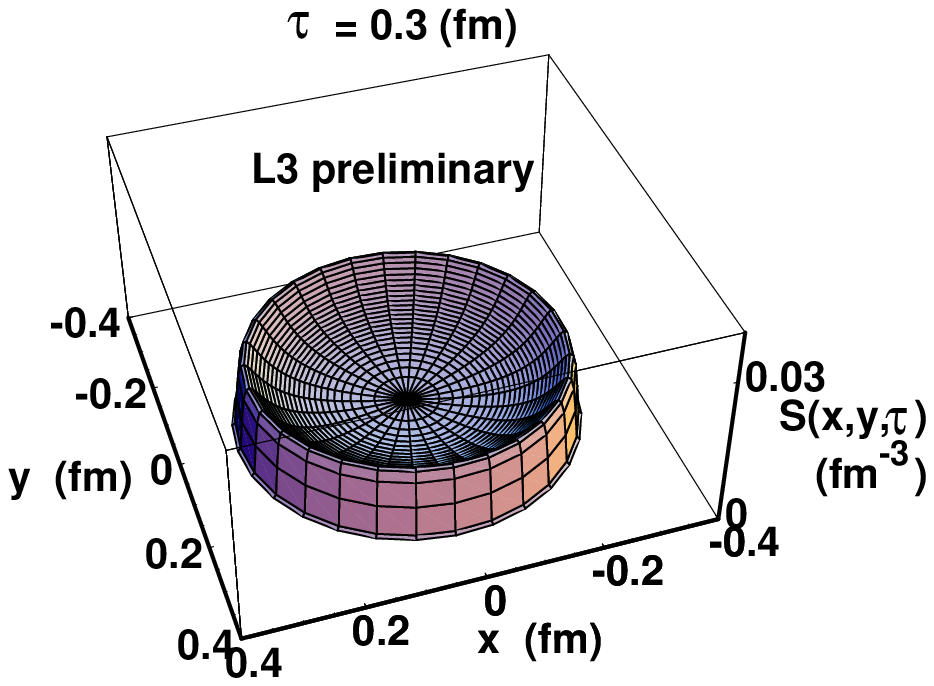}
  \caption{The transverse source function normalized to the average number of pions per event for
           various proper times.
\label{fig:xymovie}
}
\end{figure}

%The transverse profile, which follows from \Eq{eq:Sspace}, has the following form:
%\begin{equation}   \label{eq:transmovie}
%   \frac{\mathrm{d}^2 n}{\mathrm{d}\tau\mathrm{d}r}
%           = \frac{\mt^3}{\tau^3} H(\tau) \rho_1\left( k=\frac{\mt r}{\tau} \right) \;.
%\end{equation}
The transverse part of the emission function is obtained by integrating over $z$ and azimuthal angle.
%$\rho_1$ is determined from the inclusive \pt\ distribution.
%The particle production probability is proportional to the proper-time distribution $H(\tau)$.
Figure \ref{fig:xymovie} shows the transverse part of the emission function for
various proper times. Particle production starts immediately, increases rapidly and decreases slowly.
A ring-like structure, similar to the expanding, ring-like wave created by a pebble in a pond, is observed.
These pictures together form a movie that ends in about 3.5\,fm, making it the shortest movie ever made
of a process in nature.
An animated gif file covering the first  0.3\,fm
(10$^{-24}$\,sec)
%= 0--10$^{-24}$\,sec
is available~\cite{tamas_movie}.

%\enote{Add longitudinal, \ie, $z$ \vs\ $r$, ``movie''}

\section{Discussion}
%\enote{This all can probably be better explained, but must be short.}
BEC of all events as well as two- and three-jet events are observed to
be well-described by a L\'evy parametrization
incorporating strong correlations between configuration- and momentum-space.
%, whereas Gaussian and Edgeworth parametrizations fail.
%\mnote{refs}\enote{Need references.}
A L\'evy distribution arises naturally from a fractal,
or from a random walk or anomalous diffusion~\cite{metzler},
and the parton shower of the leading log
approximation of QCD is a fractal \cite{Dahlqvist:1989yc,Gustafson:1990qi,Gustafson:1991ru}.
In this case,
the L\'evy index of stability is related to the strong coupling
constant, $\as$, by\cite{alphasLevy,alphasLevy:arx}
%\mnote{check}\enote{Check formula.}
\begin{equation}
    \as = \frac{2\pi}{3} \alpha^2 \;.
\end{equation}
Assuming (generalized) local parton hadron duality \cite{Azimov:lphd,Azimov:lphd1,genlphd},
one can expect that the distribution of hadrons retains the features of the gluon distribution.
For the value of $\alpha$ found in fits of \Eq{eq:asymlevR2} we find
$\as=0.37\pm0.04$ for two-jet events,
%\mnote{check}\enote{Check values.}
%\mnote{value}\enote{Need value for systematics.}
This is a reasonable value for a scale of 1--2\,\GeV, which is where the production of hadrons takes place.
For comparison, from $\tau$ decay, $\as(m_\tau\approx1.8\,\GeV)=0.35\pm0.03$ \cite{PDG:2004}.
%\mnote{!!!!}\enote{This is a huge value if we argue that it should be at the scale of $\sqrt{s}$.
%                   On the other hand, it would seem reasonable for a scale of 1--2\,\GeV.
%                   But at this low scale, does the fractal picture still hold?
%                  }
%At first sight, these values of \as\ seem too small.
%However, we note that they are similar to the value,   $ 0.0805 \pm 0.0010$,
%found~\cite{L3:QCDphysrep}
%from a fit of the leading log formula to the \rs\ dependence of the mean multiplicity in
%$\Pep\Pem\rightarrow\mathrm{hadrons}$, while a fit using the next-to-next-to-next-to-leading order
%resulted in a value of $0.129\pm0.003$ consistent with more conventional determinations of $\as(M_\PZ)$.

It is of particular interest to point out the \mt\ dependence of the ``width'' of the source.
In \Eq{eq:levy2jetR2} the parameter associated with the width is $\Delta \tau$.
Note that it enters \Eq{eq:levy2jetR2} as $\Delta\tau Q^2/\overline{m}_\mathrm{t}$.
In a Gaussian parametrization the radius $R$ enters the parametrization as $R^2Q^2$.
Our observance that $\Delta\tau$ is independent of $\overline{m}_\mathrm{t}$ thus corresponds to
$R\propto1/\sqrt{\overline{m}_\mathrm{t}}$
and can be interpreted as confirmation of the observance~\cite{Smirnova:Nijm96,Dalen:Maha98}
of such a dependence of the Gaussian radii in 2- and 3-dimensional analyses of Z decays.
The lack of  dependence of all the parameters of \Eq{eq:levy2jetR2} on the transverse mass is in
accordance with the $\tau$-model.
 
%\mnote{!!!!}\enote{Discussion of emission function.}
%\enote{\color{BrickRed}  Can Wolfram or Tamas C. provide something for here?}
Using the BEC fit results and the $\tau$-model, the emission function of two-jet events is reconstructed.
Particle production begins immediately after collision, increases rapidly and then decreases slowly,
occuring predominantly close to the light cone.
In the transverse plane a ring-like structure expands outwards,
which is similar to the picture in hadron-hadron interactions but unlike that of heavy ion collisions.

% Acknowledgements
%%%%%%%%%%%%%%%%%%%%%%%%%%%%%%%%%%%%%%%%%%%%%%%%%%%%%%%%%%%%%%%%%%%%%%%%%%%%%%%
%
%\section*{Acknowledgments}
%We thank T. Cs\"org\H{o} for many enlightening and useful discussions.
%We wish to express our gratitude to the CERN accelerator divisions for
%the excellent performance of the LEP machine.
%We acknowledge the contributions of the engineers
%and technicians who have participated in the construction
%and maintenance of this experiment.
 
%
%%%%%%%%%%%%%%%%%%%%%%%%%%%%%%%%%%%%%%%%%%%%%%%%%%%%%%%%%%%%%%%%%%%%%%%%%%%%%%%
% The author list
%%%%%%%%%%%%%%%%%%%%%%%%%%%%%%%%%%%%%%%%%%%%%%%%%%%%%%%%%%%%%%%%%%%%%%%%%%%%%%%
%
%\newpage
%\section*{Author List}
%\input /l3/paper/namelist213.tex
\newpage
 
% Style file to use with mcite.
% Use l3style with just cite.
%\bibliographystyle{/l3/paper/biblio/l3stylem}
%\bibliographystyle{\lthreebiblio/l3stylem}
\bibliographystyle{\lthreebiblio/l3style}
\bibliography{%
%/afs/cern.ch/user/m/metzger/tex/biblio/l3,%
%/afs/cern.ch/user/m/metzger/tex/biblio/generators,%
%/afs/cern.ch/user/m/metzger/tex/biblio/jets,%
%/l3/paper/biblio/l3pubs,%
%/l3/paper/biblio/aleph,%
%/l3/paper/biblio/delphi,%
%/l3/paper/biblio/opal,%
%/l3/paper/biblio/markii,%
%/l3/paper/biblio/otherstuff,%
l3,%
generators,%
jets,%
\lthreebiblio/l3pubs,%
\lthreebiblio/aleph,%
\lthreebiblio/delphi,%
\lthreebiblio/opal,%
\lthreebiblio/markii,%
\lthreebiblio/otherstuff,%
bec,levy%
}

\hyphenation{Post-Script Sprin-ger}
\begin{thebibliography}{10}

\bibitem{Gyulassy:1979}
M. Gyulassy, S.K. Kauffmann and Lance W. Wilson,
\newblock  Phys. Rev. {\bf C20}  (1979) 2267--2292.

\bibitem{Boal:1990}
David H. Boal, Claus-Konrad Gelbke and Byron K. Jennings,
\newblock  Rev. Mod. Phys. {\bf 62}  (1990) 553--602.

\bibitem{Baym:1998}
Gordon Baym,
\newblock  Acta Phys. Pol. {\bf B29}  (1998) 1839--1884.

\bibitem{Wolfram:Zako2001}
W. Kittel,
\newblock  Acta Phys. Pol. {\bf B32}  (2001) 3927--3972.

\bibitem{Tamas:HIP2002}
T. Cs\"org\H{o},
\newblock  Heavy Ion Physics {\bf 15}  (2002) 1--80.

\bibitem{l3:construction}
L3 Collab., B. Adeva \etal,
\newblock  Nucl. Inst. Meth. {\bf A 289}  (1990) 35--102.

\bibitem{L3:Ecal_calib}
J.A.~Bakken \etal,
\newblock  Nucl. Inst. Meth. {\bf A 275}  (1989) 81--88.

\bibitem{L3:Hcal}
O.~Adriani \etal,
\newblock  Nucl. Inst. Meth. {\bf A 302}  (1991) 53--62.

\bibitem{L3:TEC}
K.~Deiters \etal,
\newblock  Nucl. Inst. Meth. {\bf A 323}  (1992) 162--168.

\bibitem{l3:SMD}
M.\ Acciarri \etal,
\newblock  Nucl. Inst. Meth. {\bf A 351}  (1994) 300--312.

\bibitem{tamas}
Tam\'as Nov\'ak,
\newblock  Ph.D. thesis, Radboud Univ.~Nijmegen, {\it in preparation}.

\bibitem{durham}
Yu.L. Dokshitzer,
\newblock  {Contribution cited in Report of the Hard QCD Working Group, Proc.
  Workshop on Jet Studies at LEP and HERA, Durham, Dec. 1990}, {J. Phys.} G17
  (1991) 1537.

\bibitem{durham2}
S. Catani \etal,
\newblock  Phys. Lett. {\bf B269}  (1991) 432--438.

\bibitem{durham3}
S. Bethke \etal,
\newblock  Nucl. Phys. {\bf B370}  (1992) 310--334.

\bibitem{L3_3D:1999}
{L3 Collab.}, M. Acciarri \etal,
\newblock  Phys. Lett. {\bf B458}  (1999) 517--528.

\bibitem{OPAL3D:2000}
{OPAL Collab.}, G. Abbiendi \etal,
\newblock  Eur. Phys. J. {\bf C16}  (2000) 423--433.

\bibitem{DELPHI2D:2000}
{DELPHI Collab.}, P. Abreu \etal,
\newblock  Phys. Lett. {\bf B471}  (2000) 460--470.

\bibitem{ALEPH:2004}
{ALEPH Collab.}, A. Heister \etal,
\newblock  Eur. Phys. J. {\bf C36}  (2004) 147--159.

\bibitem{Tamas;Lorstad:1996}
T. Cs\"org\H{o} and B. L\"orstadt,
\newblock  Phys. Rev. {\bf C54}  (1996) 1390--1403.

\bibitem{NA22q}
NA22 Collab., N.M. Agababyan \etal,
\newblock  Z. Phys. {\bf C71}  (1996) 405--414.

\bibitem{Solz:thesis}
Ron A. Solz,
\newblock  Two-Pion Correlation Measurements for 14.6A$\cdot$\GeV/$c$
  $^{28}$Si+X and 11.6A$\cdot$\GeV/$c$ $^{197}$Au+Au,
\newblock  Ph.D. thesis, Massachusetts Inst. of Technology., 1994.

\bibitem{Solz:paper}
E-802 Collab., L. Ahle \etal,
\newblock  Phys. Rev. {\bf C66}  (2002) 054906{--}1--15.

\bibitem{TASSO:1986}
{TASSO Collab.}, M. Althoff \etal,
\newblock  Z. Phys. {\bf C30}  (1986) 355--369.

\bibitem{GGLP:1960}
Gerson Goldhaber, Sulamith Goldhaber, Wonyong Lee and Abraham Pais,
\newblock  Phys. Rev. {\bf 120}  (1960) 300--312.

\bibitem{Bolz:1992hc}
J. Bolz \etal,
\newblock  Phys. Rev. {\bf D47}  (1993) 3860--3870.

\bibitem{Csorgo:1994in}
T. Cs\"org\H{o}, B. L\"{o}rstad and J. Zim\'{a}nyi,
\newblock  Z. Phys. {\bf C71}  (1996) 491--497.

\bibitem{Tamas:Moriond28}
T. Cs\"org\H{o} and S. Hegyi,
\newblock  in Proc. XXVIIIth Rencontres de Moriond, ed. {\'Etienne Aug\'e and
  J.~Tr\^an~Thanh V\^an},  (Editions Fronti\`eres, Gif-sur-Yvette, France,
  1993), p. 635.

\bibitem{Tamas:Cracow1994}
T. Cs\"org\H{o},
\newblock  in Proc. Cracow Workshop on Multiparticle Production, ed.
  {A.~Bia{\l}as \etal},  (World Scientific, Singapore, 1994), p. 175.

\bibitem{Edgeworth}
F.Y. Edgeworth,
\newblock  Trans. Cambridge Phil. Soc. {\bf 20}  (1905) 36,
\newblock  see also, \eg, Harald Cram{\'e}r, Mathematical Methods of
  Statistics, (Princeton Univ. Press, 1946).

\bibitem{Nolan}
J.P. Nolan,
\newblock  {Stable distributions: Models for Heavy Tailed Data}, 2005,
\newblock  {
  \tt{http://academic2.american.edu/$\sim$jpnolan/}\\{stable/CHAP1.PDF}}.

\bibitem{Tamas:Levy2004}
T. Cs\"org\H{o}, S. Hegyi and W.A. Zajc,
\newblock  Eur. Phys. J. {\bf C36}  (2004) 67--78.

\bibitem{Smirnova:Nijm96}
B.~L\"orstad and O.G.~Smirnova,
\newblock  in Proc. 7$^\mathrm{th}$ Int. Workshop on Multiparticle Production
  ``Correlations and Fluctuations'', ed. {R.C.~Hwa \etal},  (World Scientific,
  Singapore, 1997), p.~42.

\bibitem{Dalen:Maha98}
J.A.~van Dalen,
\newblock  in Proc. 8$^\mathrm{th}$ Int. Workshop on Multiparticle Production
  ``Correlations and Fluctuations '98: From QCD to Particle Interferometry'',
  ed. {T.~Cs\"org\H{o} \etal},  (World Scientific, Singapore, 1999), p.~37.

\bibitem{Bialas:1999}
A. Bia{\l}as and K.~Zalewski,
\newblock  Acta Phys. Pol. {\bf B30}  (1999) 359--367.

\bibitem{Bialas:2000}
A. Bia{\l}as \etal,
\newblock  Phys. Rev. {\bf D62}  (1999) 114007{--}1--7.

\bibitem{Gottfried:1972}
K. Gottfried,
\newblock  Acta Phys. Pol. {\bf B3}  (1972) 769.

\bibitem{Bjorken:SLAC73}
J.D. Bjorken,
\newblock  in Proc. Summer Inst. on Particle Physics, Vol.~1,  (SLAC-R-167,
  1973), pp. 1--34.

\bibitem{Bjorken:1973}
J.D. Bjorken,
\newblock  Phys. Rev. {\bf D7}  (1973) 282--283.

\bibitem{Gottfried:1974}
K. Gottfried,
\newblock  Phys. Rev. Lett. {\bf 32}  (1974) 957--961.

\bibitem{Low:1978}
F.E.~Low and K.~Gottfried,
\newblock  Phys. Rev. {\bf D17}  (1978) 2487--2491.

\bibitem{Bjorken:ISMD94}
J.D. Bjorken,
\newblock  in Proc. XXIV Int. Symp. on Multiparticle Dynamics, ed.
  {A.~Giovannini \etal},  (World Scientific, Singapore, 1995), p. 579.

\bibitem{Tamas;Zimanji:1990}
T. Cs\"org\H{o} and J. Zim\'anyi,
\newblock  Nucl. Phys. {\bf A517}  (1990) 588--598.

\bibitem{Yano}
F.B. Yano and S.E. Koonin,
\newblock  Phys. Lett. {\bf B78}  (1978) 556--559.

\bibitem{NA22emiss}
NA22 Collab., N.M. Agababyan \etal,
\newblock  Phys. Lett. {\bf B422}  (1998) 359--368.

\bibitem{HIemiss:Ster}
A. Ster, T. Cs\"org\H{o} and B. L\"orstad,
\newblock  Nucl. Phys. {\bf A661}  (1999) 419--422.

\bibitem{tamas_movie}
T. Nov\'ak,
\newblock  {\tt{http://www.hef.kun.nl/$\sim$novakt/movie/movie.}\\{gif}}.

\bibitem{metzler}
R. Metzler and J. Klafter,
\newblock  Phys. Rep. {\bf 339}  (2000) 1--77.

\bibitem{Dahlqvist:1989yc}
P.~Dahlqvist, B.~Andersson and G.~Gustafson,
\newblock  Nucl. Phys. {\bf B328}  (1989) 76.

\bibitem{Gustafson:1990qi}
G.~Gustafson and A.~Nilsson,
\newblock  Nucl. Phys. {\bf B355}  (1991) 106.

\bibitem{Gustafson:1991ru}
G.~Gustafson and A.~Nilsson,
\newblock  Z. Phys. {\bf C52}  (1991) 533.

\bibitem{alphasLevy}
T. Cs\"org\H{o} \etal,
\newblock  Acta Phys. Pol. {\bf B36}  (2005) 329--337.

\bibitem{alphasLevy:arx}
T. Cs\"org\H{o} \etal,
\newblock  {Bose-Einstein or HBT correlation signature of a second order QCD
  phase transition}, 2005,
\newblock  \mbox{\tt http://arXiv.org/abs/nucl-th/0512060}.

\bibitem{Azimov:lphd}
Ya.I. Azimov \etal,
\newblock  Z. Phys. {\bf C27}  (1985) 65--72.

\bibitem{Azimov:lphd1}
Ya.I. Azimov \etal,
\newblock  Z. Phys. {\bf C31}  (1986) 213--218.

\bibitem{genlphd}
L.~Van Hove and A.~Giovannini,
\newblock  Acta Phys. Pol. {\bf B19}  (1988) 931--946.

\bibitem{PDG:2004}
Particle Data Group, S. Eidelman \etal,
\newblock  Phys. Lett. {\bf B592}  (2004) 1--1109.

\end{thebibliography}

\end{document}